\documentclass[aps,twocolumn,amsfonts,amssymb,longbibliography,superscriptaddress]{revtex4-1}
\usepackage{graphicx}
\usepackage{multirow}
\usepackage{epstopdf} %converting to PDF
\usepackage{dcolumn}
\usepackage{bm}
\usepackage{bbm,bbold}
\usepackage{color}
\usepackage{xcolor, soul}
\usepackage{pifont}
\sethlcolor{green}
\usepackage{amsfonts}
\usepackage{amsmath}
\usepackage{mdframed}
\usepackage[normalem]{ulem}
\usepackage{mathrsfs}   
\usepackage[none]{hyphenat}
\usepackage{subfigure}
\usepackage{float}
\usepackage{physics}
\usepackage{booktabs}
\usepackage[colorlinks=true,citecolor=blue]{hyperref}
\hypersetup{colorlinks=true,citecolor=blue,linkcolor=blue,urlcolor=blue}

\begin{document}
	\title{Geometric Photon-drag Effect in Unconventional Magnets}	
	\author{Bristi Ghosh}
\thanks{These authors contributed equally}
\affiliation{School of Basic Sciences, Indian Institute of Technology Bhubaneswar, Argul, Jatni, Khurda, Odisha 752050, India}

\author{Vivek Pandey}
\thanks{These authors contributed equally}
\affiliation{Department of Physics, Indian Institute of Technology Bombay, Powai, Mumbai 400076, India}
\affiliation{Department of Physics, SRM University, Andhra Pradesh-522502, India.}

\author{Malay Bandyopadhyay}
\email{malay@iitbbs.ac.in}
\affiliation{School of Basic Sciences, Indian Institute of Technology Bhubaneswar, Argul, Jatni, Khurda, Odisha 752050, India}

\author{Pankaj Bhalla}
\email{pankaj.b@srmap.edu.in}
\affiliation{Department of Physics, SRM University, Andhra Pradesh-522502, India.}
\affiliation{Asian Consortium on Computational Materials Science – Global Research Centre, SRM University, AP, Andhra Pradesh 522503, India}

\author{Snehasish Nandy}
\email{snehasish@phy.nits.ac.in}
\affiliation{Department of Physics, National Institute of Technology Silchar, Assam 788010, India}

\date{\today}

\begin{abstract}
%
% The photon-drag effect is a nonlinear optical phenomenon in which the finite momentum carried by incident photons is transferred to charge carriers, generating a dc photocurrent even in centrosymmetric systems. Here, we investigate the geometric photon-drag effect in unconventional $p_x$- and $d_{x^2-y^2}$-wave altermagnets in the presence of Rashba spin-orbit coupling. Using the density-matrix formalism under finite photon-momentum excitation, we derive the nonlinear injection and shift conductivities associated with the photon-drag photogalvanic response. A central finding is that, despite the absence of inversion symmetry, the conventional photogalvanic response vanishes identically in the ($q=0$) limit upon Brillouin zone integration in both altermagnetic phases, rendering the resulting nonlinear dc photocurrent purely photon-drag in origin. The two phases exhibit qualitatively distinct responses governed by their symmetry properties: the $p$-wave magnet supports only linear injection and circular shift conductivities, whereas the $d$-wave altermagnet, which simultaneously breaks time-reversal and inversion symmetries, admits all linear and circular components. The conductivity spectra display characteristic low-frequency divergences from intraband processes, pronounced interband resonances near the Pauli-unblocked region boundaries, and bipolar peak structures arising from Fermi-surface filtering.

The photon-drag effect is a nonlinear optical phenomenon in which the finite momentum carried by incident photons is transferred to charge carriers, thereby generating a dc photocurrent even in systems with inversion symmetry. Within the density-matrix formalism in the velocity gauge, and accounting for nonvertical optical transitions driven by finite photon momentum, we derive the nonlinear injection and shift conductivities that characterize the photon-drag photogalvanic response. We investigate these photon-drag responses in unconventional $p_x$-wave and $d_{x^2-y^2}$-wave magnets in the presence of Rashba spin-orbit coupling (RSOC), highlighting the distinct roles of symmetry and quantum geometry in shaping their nonlinear optical behavior. The key finding of our work is that, despite the broken inversion symmetry, the conventional photogalvanic response vanishes identically in the $q=0$ limit after Brillouin-zone integration in both unconventional magnetic phases, resulting in a nonlinear dc photocurrent entirely photon-drag driven. The two unconventional magnets exhibit qualitatively distinct responses governed by their symmetry properties: the $p$-wave magnet supports only linear injection and circular shift conductivities, whereas the $d$-wave altermagnet, which simultaneously breaks time-reversal and inversion symmetries, admits all linear and circular components. Our findings establish altermagnets as a promising platform for controlling symmetry-selective nonlinear photocurrents and may enable applications in polarization-sensitive photodetection, nonlinear optoelectronics, and quantum geometric photogalvanics.
\end{abstract}

\maketitle

\section{Introduction}

%=================================================================
\begin{figure}[t]
    % \centering
    \includegraphics[width=\columnwidth]{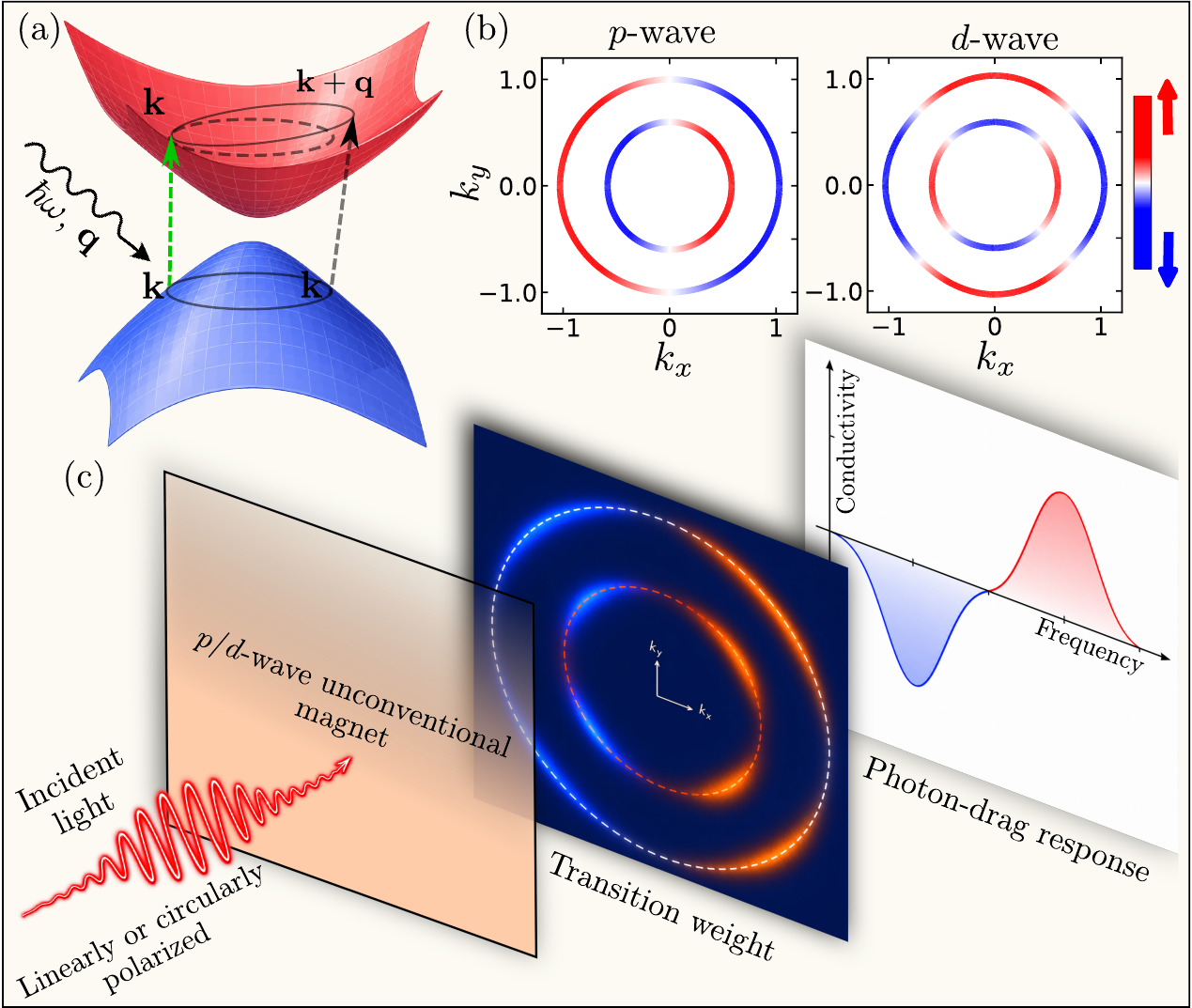}
    \caption{Schematic (a) depicts the origin of this current from vertical and nonvertical optical transitions. In the case of vertical transitions, the photon transfers only its energy $\hbar \omega$, conserving the crystal momentum ($\Delta \bm k=0$). On the contrary, nonvertical transitions involve both the energy and finite photon momentum $\hbar \bm q$, resulting in a transition from $\bm k$ to $\bm k+\bm q$. The solid and dashed circles depict constant energy surfaces corresponding to vertical and nonvertical transitions, respectively. Panel (b) represents the Fermi surface (constant energy contour at $\varepsilon=\varepsilon_F$) of the $p$- and $d$-wave altermagnets. The red and blue contours correspond to the two spin-split bands. Here, the $p$-wave exhibits a nearly circular Fermi surface, whereas the $d$-wave shows a characteristic fourfold anisotropic Fermi surface due to the $d_{x^2-y^2}$-wave exchange field. Panel (c) represents the light of energy $\hbar \omega$ and momentum $\hbar \bm q$ incident on the altermagnet sample in the presence of SOC, which generates a nonzero photocurrent response arising from the nonvertical transition in the centrosymmetric materials.}
    \label{fig:Schematic}
\end{figure}
%=================================================================

Unconventional magnets, characterized by momentum-dependent and symmetry-enforced spin-split electronic band structures, have recently emerged as a fundamentally new class of magnetic quantum materials~\cite{libor_prx2022_2, libor_prx2022, liu_np2025, Yuan_prb2026, hayami_jpsj2019, liu_npjqm2024, Yuan_prb2020, Igor_pnas2021}. Among them, altermagnets constitute a particularly remarkable family, where neighboring collinear magnetic moments alternate in sign according to a nonrelativistic spin-group symmetry rather than conventional spin-orbit coupling mechanisms~\cite{libor_prx2022_2, Ahn_prb2019, Mazin2024, Cheong2025}. Unlike conventional collinear antiferromagnets, altermagnets explicitly break time-reversal symmetry ($\mathcal{T}$), leading to pronounced momentum-space ($\mathbf{k}$-dependent) spin polarization and spin-split electronic bands despite possessing zero net magnetization~\cite{hayami_jpsj2019, Jungwirth2022}. More generally, unconventional magnetic orders can be classified according to the parity of the magnetic order parameter. Even-parity magnetic orders, exemplified by $d$-wave altermagnets, break $\mathcal{T}$ while preserving inversion symmetry ($\mathcal{P}$), whereas odd-parity magnetic orders, such as $p$-wave magnets, preserve $\mathcal{T}$ but break $\mathcal{P}$~\cite{hellenes2023,Linder_2024_PRL,Zhu_2023}. A defining characteristic of these unconventional magnetic phases is that their momentum-dependent spin splitting is not uniform throughout the Brillouin zone but vanishes at symmetry-protected nodal points dictated by the underlying magnetic form factor~\cite{libor_prx2022, Yang2025}. 

The unique spin texture of unconventional magnetic systems gives rise to a plethora of intriguing electrical and optical responses in connection with quantum geometry in these systems. In the case of electrical responses, it has been shown that the linear and second-order anomalous Hall effects originating from the Berry monopole and Berry dipole respectively, are symmetry forbidden despite the presence of sizable momentum-dependent spin splitting~\cite{Korrapati_prb2025, korrapati2025, Sodemann2015, wang_n2023, Gao_science2023, Arnob_2026}, leaving the third-order responses as the leading-order governed by the higher-order pole of Berry curvature~\cite{Fang_prl2024}. In the optical regime, the symmetries of unconventional magnets also restrict second-order optical phenomena, such as second-harmonic generation and the photogalvanic effect, in the absence of external symmetry-breaking fields~\cite{Ezawa_2025,Werner_2024}, making the realization of a finite second-order optical response without such external fields particularly challenging.

Recently, a unique contribution to the second-order photocurrent driven by the quantum geometry, namely, the photon-drag effect, originating from the transfer of the finite momentum of light to Bloch electrons, has attracted considerable attention in recent theoretical and experimental studies~\cite{xie_pnas2025,Geng_2025,Shi_prl2021,Qu_2025,lu2025giant}. Unlike the conventional photogalvanic effect, the photon-drag effect can generate second-order dc current even in centrosymmetric materials~\cite{Geng_2025,Shi_prl2021}. Although the photon-drag effect has been extensively investigated in centrosymmetric systems ~\cite{xie_pnas2025,Shi_prl2021}, its behavior in unconventional magnetic systems remains largely unexplored. In particular, it remains unclear whether the photon-drag response in unconventional magnets is primarily governed by the presence or absence of inversion symmetry, or whether additional ingredients, such as spin-orbit coupling and the underlying magnetic symmetry, are essential. These naturally raise several fundamental questions: can unconventional magnetic order alone induce a finite dc photocurrent driven by the photon-drag effect without lowering the crystal symmetry or applying external fields? More importantly, can the photon-drag effect serve as a direct fingerprint of unconventional magnetism, enabling the distinction between even-parity and odd-parity magnetic orders through their underlying quantum geometry?

In this work, we investigate the photon-drag response within the density-matrix formalism in the velocity gauge in odd-parity unconventional $p$-wave magnets and even-parity $d$-wave altermagnets to address these questions. Unlike conventional vertical optical transitions, which conserve crystal momentum, photon-drag processes involve the transfer of finite photon momentum $\mathbf{q}$ to the electronic system, thereby enabling nonvertical transitions between momentum-shifted electronic states and generating a dc photocurrent [Fig.~\ref{fig:Schematic}(a)]. Despite their distinct momentum-dependent spin-split Fermi surfaces as illustrated in Fig.~\ref{fig:Schematic}(b), both magnetic phases exhibit a characteristic bipolar structure in the photon-drag response, with the two peaks originating from optical excitations associated with the two spin-split Fermi surfaces [Fig.~\ref{fig:Schematic}(c)]. We show that, while neither the $p$-wave nor the $d$-wave magnet exhibits the conventional photogalvanic effect irrespective of the presence of Rashba spin-orbit coupling (RSOC), the combined effect of unconventional magnetic order and RSOC activates a finite photon-drag response in the absence of external fields, whereas no photon-drag response is found without RSOC. By systematically analyzing the injection and shift current contributions, we demonstrate that the symmetry and quantum-geometric characteristics of the odd- and even-parity magnetic phases generate distinct nonlinear optical signatures, establishing the photon-drag effect as a direct fingerprint of unconventional magnetism. 
% \textcolor{blue}{The physical picture underlying the photon-drag process, together with the characteristic electronic structures of the two magnetic phases, is schematically illustrated in Fig. 1. Unlike conventional vertical optical transitions, which conserve crystal momentum, photon-drag processes involve the transfer of the finite photon momentum $\hbar \mathbf{q}$ to the electronic system, thereby enabling nonvertical transitions between momentum-shifted electronic states [Fig. 1(a)]. This finite-momentum transfer provides the microscopic origin of the dc photocurrent investigated here. The resulting response is intimately connected to the underlying Fermi-surface geometry: as illustrated in Fig. 1(b), the $p$-wave magnet exhibits an approximately circular Fermi surface, whereas the $d_{x^2 −y^2}$ wave altermagnet displays a pronounced fourfold anisotropy arising from its momentum-dependent exchange splitting. Figure 1(c) further depicts the optical excitation geometry, in which light carrying energy $\hbar \omega$ and momentum $\hbar q$ drives these nonvertical transitions and generates a finite photon-drag photocurrent. Together, these schematics highlight the central physical ingredients of the response—the finite momentum of light, the symmetry-dependent Fermi-surface structure, and the quantum geometry of the electronic states—and provide a visual overview of the mechanism underlying the distinct nonlinear optical responses of the two unconventional magnetic phases.} 
Specifically, the $p$-wave magnet, which preserves $\mathcal{T}$ while breaking $\mathcal{P}$, supports only linear injection and circular shift conductivities, whereas the $d$-wave altermagnet, which simultaneously breaks both symmetries, admits all linear and circular components of the injection and shift responses. The different symmetry properties of the two magnetic phases consequently give rise to different symmetry-allowed components of the photon-drag response. These findings, which are experimentally accessible in several unconventional magnetic materials such as CeNiAsO~\cite{hellenes2023}, NiI$_2$~\cite{song2025}, RuO$_2$~\cite{Ahn_prb2019} and MnF$_2$~\cite{Bhowal_2024}, establish the photon-drag effect as a sensitive probe of quantum geometry in unconventional magnets. Moreover, it provides a promising approach for distinguishing different unconventional magnetic orders, with potential applications in polarization-sensitive photodetection, optical rectification, and next-generation optoelectronic devices.

This paper is organized as follows. Section~\ref{sec:model and method} presents the quantum kinetic formalism for the photon-drag effect, followed by the $p$- and $d$-wave unconventional magnetic model Hamiltonians in Sec.~\ref{subsec:model}. The numerical results are presented in Sec.~\ref{sec:results}, where the photon-drag responses of the two altermagnetic phases are discussed separately. In particular, Sec.~\ref{sec:$p_x$} examines the injection and shift conductivities of the p-wave magnet, while Sec.~\ref{sec:$d_wave$} presents the corresponding results for the $d$-wave altermagnet, enabling a direct comparison of their symmetry-dependent nonlinear optical responses. Finally, Sec.~\ref{sec: conclusion} summarizes the main conclusions of this work.

% Remarkably, although the photon-drag effect is symmetry allowed even in centrosymmetric materials, we find no finite response in the centrosymmetric d-wave magnet in the absence of Rashba spin-orbit coupling. A finite photon drag response emerges only when Rashba spin-orbit coupling is incorporated. By systematically analyzing the injection and shift current contributions, we show that the symmetry and quantum-geometric characteristics of these two classes of magnetic order produce distinct nonlinear optical signatures, establishing the photon drag effect as a direct fingerprint of unconventional magnetism.

%
%

\section{Theoretical Framework For Photon-Drag Effect}\label{sec:model and method}

%\subsection{Method}\label{subsection:method}
%
The photon-drag effect arises from the transfer of finite photon momentum ($\textbf{q}$) to Bloch electrons, enabling a dc photocurrent even in systems where the conventional photogalvanic effect vanishes. In order to derive the general expression of second-order optical currents incorporating the photon momentum of the incident wave, we employ the quantum Liouville equation~\cite{Culcer_2017, Bhalla_prb2023}, which governs the time evolution of the single-particle density matrix $\rho(\mathbf{k},t)$,
\begin{align}
\label{QLE}
    \frac{\partial\rho(\mathbf{k},t)}{\partial t}+ \frac{i}{\hbar}[\mathcal{H}(\mathbf{k},t), \rho(\mathbf{k},t)]=0,
\end{align}
where $\mathcal{H}(\mathbf{k},t)$ represents the total Hamiltonian of the system. In the presence of the electromagnetic field $\mathbf{E}(\mathbf{r}, t) =\sum_{\xi=\pm} \mathbf{E}(\xi \mathbf{q}, \xi\omega)e^{i\xi(\mathbf{q}\cdot\mathbf{r} - \omega t)} $ with frequency $\omega$ and photon momentum $\mathbf{q}$, the unperturbed Hamiltonian $\mathcal{H}_0$ is modified under the velocity gauge through the minimal-coupling substitution $\mathbf{k}\rightarrow\mathbf{k}+\frac{e\mathbf{A}}{\hbar}$, giving the perturbed Hamiltonian as 
$\mathcal{H}(\mathbf{k},t)=\mathcal{H}_0( {{\textbf{k}}}+\frac {e\mathbf{A}(\mathbf{r},t)}{\hbar} )$. Here, $\mathbf{A}(\mathbf{r},t)$ is a vector potential satisfying $\mathbf{E}(\mathbf{r},t)=-\frac{\partial \mathbf{A}(\mathbf{r},t)}{\partial t}$.

The presence of the electromagnetic field introduces a correction to the Hamiltonian, which can be written as $\mathcal{H}(\mathbf{k},t) \approx \mathcal{H}_0({\mathbf{k}}) + \mathcal{H}_{Int}(\mathbf{r}, t)$,  where $\mathcal{H}_0({\textbf{k}})$ is the unperturbed Hamiltonian, and $\mathcal{H}_{Int}(\mathbf{r}, t)$ indicates the interaction between the optical field and the system. Retaining terms up to second order in the vector potential, the interaction Hamiltonian can be decomposed as $\mathcal{H}_{Int}(\mathbf{r}, t)= H^{\prime}(\mathbf{r}, t) + H^{\prime \prime}(\mathbf{r}, t)$, where $H^{\prime}(\mathbf{r}, t)$ and $ H^{\prime\prime}(\mathbf{r}, t)$ depend linearly and quadratically on the vector potential (electric field), respectively, as 
\begin{align}
    H^{\prime}(\mathbf{r}, t)=e A_a(\mathbf{r},t)v^a,
    \label{Eq.2}
    \end{align}
\begin{align}
    H^{\prime \prime}(\mathbf{r}, t)=\frac{e^2}{2} A_a(\mathbf{r},t) A_b(\mathbf{r},t) v^{ab},
        \label{Eq.3}
\end{align}
where $v^a=\frac{1}{\hbar}\frac{\partial {\color{black}\mathcal{H}_0}}{ \partial k_a}$ and $v^{ab}=\frac{1}{\hbar^2}\frac{\partial^2 {\color{black}\mathcal{H}_0}}{ \partial k_a \partial k_b}$.
In the presence of the photon-drag effect, an electron with a wave vector $\bf k$ undergoes a transition from $\ket{n, \bf k -\frac{\bf q}{2}}$ to $\ket{m, \bf k +\frac{\bf q}{2}}$ and vice versa, where $m,n$ are the band indices.
Now, the interaction Hamiltonian in the band basis has the following form,
\begin{align}
    H^{\prime}_{mn}(\mathbf{k}, t)=e\sum_{\xi=\pm}A_a( \xi \mathbf{q},\xi\omega)v^a_{m,\textbf{k}^\xi_{+}; n,\textbf{k}^\xi_{-}}e^{-i\xi\omega t}.
\end{align}
%{\color{red}The notation is bit confusing here. Earlier, the direction for velocity is written as subscript and now it is written as superscript. Better to be consistent.}
\begin{align}
    H^{\prime \prime}_{mn}(\mathbf{k}, t)=&\frac{e^2}{2}\sum_{\xi_1, \xi_2=\pm}A_a(\xi_1 \mathbf{q},\xi_1\omega) A_b(\xi_2 \mathbf{q},\xi_2\omega)\nonumber \\&\times v^{ab}_{m,\textbf{k}^{\xi_{1}+\xi_2}_{+}; n,\textbf{k}^{\xi_{1}+\xi_2}_{-}}e^{-i(\xi_1+\xi_2)\omega t}.
\end{align}
where the first-order velocity matrix ($v^{a}$) in the band basis representation is written as $v^a_{m,\textbf{k}^\xi_{+}; n,\textbf{k}^\xi_{-}}={1}/{\hbar }\bra{m,\mathbf{k}+ \xi\frac{\mathbf{q}}{2}}\frac{\partial \mathcal{H}_0}{\partial k_a}\ket{ n,\bf k- \xi\frac{\mathbf{q} }{2}}$,  and $A_a(\xi\mathbf{q},\xi\omega)$ is the strength of the gauge potential, defined as, $A_a(\xi \mathbf{q},\xi\omega)=-i E_a(\xi \mathbf{q},\xi \omega)/\xi \omega$. For photon-drag photogalvanic effect we have to take $\xi_{1}=-\xi_{2}=\xi$.\\

Within the relaxation-time approximation, the quantum Liouville equation given in Eq.~(\ref{QLE}) takes the form,
\begin{align}
    \frac{\partial\rho}{\partial t}+ \frac{i}{\hbar}[\mathcal{H}_0+\mathcal{H}_{Int}, \rho]+\frac{\rho-\rho^{(0)}}{\tau}=0,
\end{align}
where the last term describes the relaxation of the nonequilibrium distribution toward the equilibrium density matrix $\rho^{(0)}$, and $\tau$ denotes the phenomenological relaxation time. In general, $\tau$ incorporates the effects of impurity scattering, electron--phonon interactions, and other dissipative processes, and may depend on momentum, frequency, and temperature~\cite{Bristi_2024}. We solve the kinetic equation perturbatively by expanding the single-particle density matrix in powers of the electric field as: $ \rho = \rho^{(0)}+\rho^{(1)}+\rho^{(2)}+....$, where $\rho^{(0)}$ is the equilibrium or unperturbed part, and  $\rho^{(1)}$ and $\rho^{(2)}$ show the linear and quadratic dependence on the electric field under the influence of the electromagnetic wave, respectively.\\ 
The dc photocurrent is evaluated from the expectation value of the current operator $\textbf{j}^{(2)}(t)=\text{Tr}[{\hat{\textbf{J}}}\rho]$, where the current operator is defined as $\hat{\textbf{J}}=-\frac{\delta \mathcal{H}}{\delta \textbf{A}}$. Now, for our case, using Eqs. (\ref{Eq.2}) and (\ref{Eq.3}), the current operator can be written as the sum of two terms, which describe
\begin{align}
    \hat{J}^{(0)}_{a}=- e v^a, \,\,
    \hat{J}^{(1)}_{a}=-e^2 A_b(\mathbf{r},t) v^{ab}.
\end{align}
So, the second-order nonlinear dc photocurrent is given by
\begin{align}
    j^{(2)}_{a}(t)&= \text{Tr}[\hat{J}^{(1)}_{a}\,\rho^{(1)}+\hat{J}^{(0)}_{a}\,\rho^{(2)}]\\ \nonumber
    &=-\text{Tr}[e^2 A_{b}{ v}^{ab}\,\rho^{(1)}+e  v^{a} \rho^{(2)}],
\end{align}
where the first term shows the interplay between the linear density matrix and the nonlinear velocity, and the second term captures the nonlinear correction to the density matrix. The second-order photocurrent can be expressed in terms of the photon-drag conductivity tensor as,
\begin{align}
    j^{2}_{a}(0)=\sigma^{abc}(\omega,-\omega)E^b(\omega)E^c(-\omega).  
\end{align}

This photon-drag conductivity arises from three distinct channels, namely the injection, shift, and Fermi-surface photocurrent contributions, denoted by $\sigma^{abc}_{\mathrm{Inj}}$, $\sigma^{abc}_{\mathrm{Sh}}$, and $\sigma^{abc}_{\mathrm{FS}}$, respectively~\cite{xie_pnas2025}. Accordingly, the total nonlinear photoconductivity can be decomposed into three contributions:
\begin{align}
    \sigma^{abc}=\sigma^{abc}_{\text{Inj}} + \sigma^{abc}_{\text{Sh}} + \sigma^{abc}_{\text{FS}}.
\end{align} %here we have focused in the injection and shift current contribution to the photocurrent. 
Physically, the shift current corresponds to the real-space shift during the optical transition, in contrast the injection current comes from the asymmetric carrier population, and the Fermi surface contribution is associated with the intraband carrier dynamics near the Fermi level and becomes important in metallic or semimetallic systems. In the present work, we have focused on the injection and shift currents, as they capture the interband nonlinear optical response and reveal the underlying geometric nature of the photon-drag photogalvanic effect. %Here, conductivity arises from the expression $j^a=\sigma^{abc} E_b(\omega)E_c(-\omega)$.
The general expressions for the injection and shift current are expressed as
\begin{widetext}
%
% \begin{align}\label{eq:gen_inj}
%    \sigma^{abc}_{\text{Inj}}=-\frac{2 \pi e^3}{\hbar^2 \omega^2 \eta}\sum_{n,m} f_{m,\textbf{k}^\xi_{+}; n,\textbf{k}^\xi_{-}} [v^a_{{m,\textbf{k}^\xi_{+}};{m,\textbf{k}^\xi_{+}}}-v^a_{{n,\textbf{k}^\xi_{-}};{n,\textbf{k}^\xi_{-}}}] v^b_{{m,\textbf{k}^\xi_{+}};{n,\textbf{k}^\xi_{-}}}v^c_{{n,\textbf{k}^\xi_{-}};{m,\textbf{k}^\xi_{+}}}\delta(\omega_{{m,\textbf{k}^\xi_{+}};{n,\textbf{k}^\xi_{-}}}-\omega),
% \end{align}
% %
% \begin{align}\label{eq:gen_shift}
% \sigma_{\text{Sh}}^{abc}= \frac{i\pi e^3}{\hbar^2 \omega^2}
% \sum_{n,m}f_{m,\textbf{k}^\xi_{+}; n,\textbf{k}^\xi_{-}}\left[v^b_{{n,\textbf{k}^\xi_{-}};{m,\textbf{k}^\xi_{+}}}
% \left(v^c_{{m,\textbf{k}^\xi_{+}};{n,\textbf{k}^\xi_{-}}}
% \right)_{; k_a}-v^c_{{n,\textbf{k}^\xi_{-}};{m,\textbf{k}^\xi_{+}}}
% \left(v^b_{{m,\textbf{k}^\xi_{+}};{n,\textbf{k}^\xi_{-}}}
% \right)_{; k_a}\right] \delta(\omega_{{m,\textbf{k}^\xi_{+}};{n,\textbf{k}^\xi_{-}}}-\omega).
% \end{align}
%
%
% \begin{align}
% \sigma_{\text{F}}^{abc}
% &= \frac{e^3}{\hbar^2 \omega^2}\sum_{n,m}\partial_{k_a} F_{nq^\xi_{+},mq^\xi_{-}}
% \frac{v^b_{nq^\xi_{+},mq^\xi_{-}} v^c_{mq^\xi_{-},nq^\xi_{+}}}{\omega_{nq^\xi_{+},mq^\xi_{-}} - \omega}.
% \end{align}
\begin{align}\label{eq:gen_inj}
   \sigma^{abc}_{\text{Inj}}=-\frac{2 \pi e^3 \tau}{\hbar^2 \omega^2 }\sum_{n,m} f_{m,\textbf{k}_{+}; n,\textbf{k}_{-}} [v^a_{{m,\textbf{k}_{+}};{m,\textbf{k}_{+}}}-v^a_{{n,\textbf{k}_{-}};{n,\textbf{k}_{-}}}]\mathbf{G}^{bc}_{{m,\textbf{k}_{+}};{n,\textbf{k}_{-}}}\delta(\omega_{{m,\textbf{k}_{+}};{n,\textbf{k}_{-}}}-\omega),
\end{align}
\begin{align}\label{eq:gen_shift}
\sigma_{\text{Sh}}^{abc}= \frac{i\pi e^3}{\hbar^2 \omega^2}
\sum_{n,m}f_{m,\textbf{k}_{+}; n,\textbf{k}_{-}}\left[\mathcal{C}^{abc}_{{m,\textbf{k}_{+}};{n,\textbf{k}_{-}}}-\mathcal{C}^{acb}_{{m,\textbf{k}_{+}};{n,\textbf{k}_{-}}}\right]\delta(\omega_{{m,\textbf{k}_{+}};{n,\textbf{k}_{-}}}-\omega).
\end{align}   
\end{widetext}
Here, $\omega_{{m,\textbf{k}_{+}};{n,\textbf{k}_{-}}}=\omega_{{m,\textbf{k}+\frac{\textbf{q}}{2}}}-\omega_{{n,\textbf{k}-\frac{\textbf{q}}{2}}}$ is the energy difference between the bands $m$ and $n$ in the presence of the photon-drag effect and the Fermi function difference is defined as $f_{m,\textbf{k}_{+}; n,\textbf{k}_{-}}=f_{{m,\textbf{k}+\frac{\textbf{q} }{2}}}-f_{{n,\textbf{k}-\frac{\textbf{q}}{2}}}$. The geometric quantity $\mathbf{G}$ and $\mathcal{C}$ are defined as 
\begin{align}
       \mathbf{G}^{bc}_{{m,\textbf{k}_{+}};{n,\textbf{k}_{-}}}&=v^b_{{m,\textbf{k}_{+}};{n,\textbf{k}_{-}}}v^c_{{n,\textbf{k}_{-}};{m,\textbf{k}_{+}}}\\
   \mathcal{C}^{abc}_{{m,\textbf{k}_{+}};{n,\textbf{k}_{-}}}&=v^b_{{n,\textbf{k}_{-}};{m,\textbf{k}_{+}}}
    \left(v^c_{{m,\textbf{k}_{+}};{n,\textbf{k}_{-}}}
    \right)_{; k_a},
\end{align}
where the covariant derivative of the velocity matrix is defined as,
\begin{equation}
    (v^c_{mn})_{; k_a}=w^{ac}_{mn}-ir^{a}_{mn}\Delta^{c}_{mn}+\sum_{p\neq m,n}\bigg(\frac{v^{a}_{mp} v^{c}_{pn}}{\omega_{mp}}-\frac{v^{c}_{mp} v^{a}_{pn}}{\omega_{pn}}\bigg)
\end{equation}
with $w^{ac}_{mn}=\hbar^{-1}\bra{m} \partial_a \partial_c \mathcal{H}_0 \ket{n}$, $r^a_{mn}= \bra{m} i\partial_a \ket{n}$ and $\Delta^{c}_{mn}=v^{c}_{mm}-v^{c}_{nn}$.
%
%To facilitate discussion, we define momentum notation as $\hat{n} \equiv \mathbf{n}k + \frac{\mathbf{q}}{2}, \space \hat{m} \equiv \mathbf{m}k - \frac{\mathbf{q}}{2}.$
%
To make a connection with the quantum geometric tensor, we would like to point out that in the limit $\bm q\to0$~\cite{xie_pnas2025, Pandey_njp2026} and moving $\omega^{-2}$ as $\omega^{-2}_{mn}$ inside the integral, the product of velocity matrix elements, $\mathbf{G}^{bc}_{{m,\textbf{k}_{+}};{n,\textbf{k}_{-}}}$ divided by $\omega^{2}_{mn}$ 
reduces to the quantum geometric tensor, whose real and imaginary parts correspond to the quantum metric and Berry curvature, respectively. Likewise, the quantity appearing in the shift conductivity, $\mathcal{C}^{abc}_{{m,\textbf{k}_{+}};{n,\textbf{k}_{-}}}$ divided by $\omega^{2}_{mn}$ reduces to the familiar quantum geometric connection in the same limit. Consequently, Eqs.~(\ref{eq:gen_inj}) and (\ref{eq:gen_shift}) recover the well-established expressions for the conventional injection and shift currents, respectively~\cite{Ahn_2020,Bristi_2026}.
\section{Model}\label{subsec:model}
Unconventional magnets are characterized by momentum-dependent spin splitting that preserves zero net magnetization while lifting spin degeneracy in a symmetry-protected manner~\cite{Ezawa_prb2025, libor_prx2022, Mazin_prx2022, gao_nsr2025, calixto_arxiv2026, ezawa_prb2025_2, Ezawa_prb2026}. In this section, we consider a minimal effective $\mathbf{k}\cdot\mathbf{p}$ model to describe two representative classes of such systems, namely $p$-wave magnets and $d$-wave altermagnets, in the presence of RSOC. The inclusion of RSOC is physically well motivated, as it naturally arises when an unconventional magnetic layer is grown on a substrate~\cite{Denis_sa2026}. Within this framework, the low-energy $\mathbf{k}\cdot\mathbf{p}$ Hamiltonian of these systems can be written as~\cite{Ezawa_prb2025}
\begin{align}
    \mathcal{H}(\bm k)=\mathcal{H}_{\text{KE}}+\mathcal{H}_{\text{RSOC}}+\mathcal{H}_{X}.
\end{align}
% 

%=================================================================
%=================================================================
\begin{table}[b]
    \caption{Symmetry properties of the $p$-wave magnet and $d$-wave altermagnet. A check mark (\ding{51}) indicates the symmetry is preserved, while a cross (\ding{55}) indicates it is broken.}
    \label{tab:symmetry}
    \begin{ruledtabular}
      \begin{tabular}{ccc}
         Symmetry & $p$-wave & $d$-wave \\
         \hline
        Time-reversal symmetry ($\mathcal{T}$) 
        & \ding{51} 
        & \ding{55}  \\
        Inversion symmetry ($\mathcal{P}$) 
        & \ding{55} & \ding{55} \\
        Mirror symmetry ($\mathcal{M}_x$) 
        & \ding{51} & \ding{55} \\
        Mirror symmetry ($\mathcal{M}_y$) 
        & \ding{55}
        &  \ding{55} \\
         $C_{4z}\mathcal{T}$ 
        & \ding{55}
        &  \ding{51} \\
        $\mathcal{M}_x\mathcal{T}$ 
        & \ding{51}
        &  \ding{55} \\
    \end{tabular}        
    \end{ruledtabular}
\end{table}
%=================================================================
%=================================================================

The first term represents the kinetic energy, $\mathcal{H}_{\text{KE}}=t k^2$, where $t=\frac{\hbar^2}{2m}$, and $m$ is the mass associated with an electron. The second term $\mathcal{H}_{\text{RSOC}}=\lambda(\bm k \times \sigma)_z=\lambda(k_x\sigma_y-k_y\sigma_x)$ 
, describes the Rashba spin–orbit coupling arising from structural inversion asymmetry, $\lambda$ denoting its strength and $\sigma_i$ are the Pauli matrices.
The third term, $\mathcal{H}_{X}$, captures the unconventional magnetic exchange field and encodes a momentum-dependent spin-splitting form factor that characterizes the underlying magnetic order. This form factor differentiates various unconventional magnets and takes distinct forms for $p$-wave and $d$-wave altermagnets. In this work, we specifically focus on $p_x$ and $d_{x^2-y^2}$-wave symmetries, which represent prototypical odd- and even-parity unconventional states, respectively. The lattice constant is set to unity without loss of generality. The exchange term $\mathcal{H}_X$ for the $p_x$-wave and $d_{x^2-y^2}$-wave symmetries is defined as,
\begin{align}
H_{p_x} &=J~k_{x}\sigma_{z}, \nonumber \\
H_{d_{x^2-y^2}} &=J~(k^2_{x}-k^2_{y})\sigma_{z},
\end{align}

\noindent where $J$ is defined as the strength of the unconventional magnetic order.

The energy dispersion for $p_x$- and $d_{x^2-y^2}$-wave unconventional magnets are given by
\begin{align}\label{Eq: dispersion}
    \mathcal{E}_{p_x}^{\pm} &= t~k^2 \pm \sqrt{J^2 k_{x}^2+\lambda^2 k^2}, \nonumber \\
    \mathcal{E}_{d_{x^2-y^2}}^{\pm} &= t~k^2 \pm \sqrt{J^2 (k_{x}^2-k_y^2)^2+\lambda^2 k^2},
\end{align}
\noindent
where $\pm$ denotes the conduction and valence bands, respectively. The band dispersion of the $p_x$-wave magnet (see Fig.~\ref{fig:Schematic}(b)) exhibits two nodes along the $k_y$ direction, with spin splitting primarily occurring along $k_x$. In contrast, the $d_{x^2-y^2}$-wave altermagnet shows four symmetry-protected nodes in the two-dimensional Brillouin zone (see Fig.~\ref{fig:Schematic}(b)), reflecting its even-parity nature. In the presence of RSOC, the $p_x$-wave magnet preserves $\mathcal{T}$ and mirror symmetry ($\mathcal{M}_x$), while spatial inversion symmetry is broken. On the other hand, the $d_{x^2-y^2}$-wave altermagnet breaks both inversion and time-reversal symmetries but preserves combined symmetries such as $C_{4z} \mathcal{T}$ and $M_{x} \mathcal{T}$, where $C_{4z}$ denotes a fourfold rotation about the $z$-axis, as summarized in Table \ref{tab:symmetry}. Consequently, it supports a finite, momentum-dependent Berry curvature unlike the $p_x$-wave magnet. 

\begin{figure*}[t]
    \includegraphics[width=0.82\textwidth]{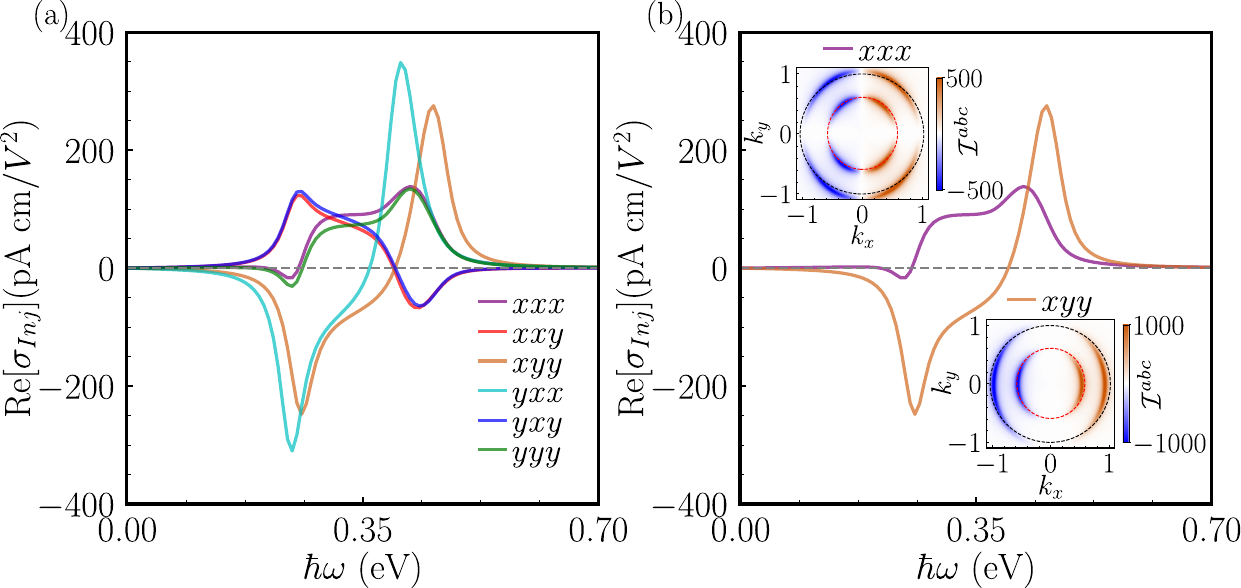}
    \caption{\textbf{Photon-drag injection response of the $p_x$-wave magnet with Rashba spin-orbit coupling:} (a) Frequency dependence of the real part of the symmetry-allowed injection conductivity. Owing to the simultaneous presence of time-reversal symmetry and broken inversion symmetry in the $p$-wave magnet, only the real part of the injection conductivity (linear injection) contributes to the photon-drag response. (b) Frequency dependence of $\mathrm{Re}[\sigma_{xxx}]$ and $\mathrm{Re}[\sigma_{xyy}]$, highlighting their distinct spectral structures and peak positions. The upper and lower insets show the corresponding geometric transition weight $\mathcal{I}^{abc}$ in the $k_x$--$k_y$ plane for the $xxx$ and $xyy$ components, respectively. Both  $\mathcal{I}^{abc}$ exhibit two dominant contributions associated with the two Fermi surfaces(red and black dashed lines). However, their different momentum-space structures result in distinct spectral contributions, giving rise to the different peak positions observed in the frequency dependence of the two conductivity components. The model parameters used are ($t$, $J$, $\lambda$, $\tau$, $T$, $q_x$, $q_y$) = ($0.5~\mathrm{eV.nm}^2$, $0.1~\mathrm{eV.nm}$, $0.2~\mathrm{eV.nm}$, $33~\mathrm{fs}$, $10~\mathrm{K}$, $0.01~\mathrm{nm}^{-1}$, $0.01~\mathrm{nm}^{-1}$).}
    \label{fig:p_wave_real_inj}
\end{figure*}
%=================================================================

%-------------------------------------
\section{Results and Discussion} \label{sec:results}
%~~~~~~~~~~~~~~~~~~~~~~~~~~~~~~~~~~~~~
In this section, we investigate the photon-drag photogalvanic response in $p_x$- and $d_{x^2-y^2}$-wave unconventional magnets, both in the presence and absence of RSOC, with an emphasis on how their distinct symmetry properties govern the resulting nonlinear photoresponse. Owing to the different forms of the exchange term $\mathcal{H}_X$, $p_x$- and $d_{x^2-y^2}$-wave systems exhibit fundamentally different symmetry constraints, which are directly reflected in their photon-drag conductivity tensors. We analyze both injection and shift current contributions to the photon-drag effect and examine their frequency dependence, comparing cases with and without RSOC to identify clear optical signatures associated with the underlying band dispersion and quantum geometric properties. Furthermore, we discuss how these responses depend on the polarization of the incident light, providing additional insight into symmetry-resolved nonlinear optical behavior.

In the absence of RSOC, both the $p_x$- and $d_{x^2-y^2}$-wave magnets possess momentum-independent eigenstates, although their momentum-dependent exchange splittings and symmetry properties are distinct. This follows from the block diagonal form of the Hamiltonian in the spin basis, which does not induce any momentum-dependent mixing between the eigenstates, leading to vanishing interband velocity matrix elements (like $v^b_{{m,\textbf{k}_{+}};{n,\textbf{k}_{-}}}$) and, correspondingly, vanishing quantum geometric quantities, $\mathbf{G}^{bc}_{{m,\textbf{k}_{+}};{n,\textbf{k}_{-}}} $ and $ \mathcal{C}^{abc}_{{m,\textbf{k}_{+}};{n,\textbf{k}_{-}}}$. As a result, all conductivity contributions contributing to both the conventional photogalvanic and finite-momentum photon-drag responses are suppressed in the absence of RSOC.
%, leading to different symmetry-allowed nonlinear responses. The p-wave phase breaks inversion symmetry while preserving time-reversal symmetry, allowing circular injection and linear shift currents for the conventional photogalvanic effect ($\textbf{q}=0$) and linear injection and circular shift currents in the photon-drag regime. In contrast, the d-wave phase preserves inversion symmetry but breaks time-reversal symmetry, thereby forbidding the conventional photogalvanic effect, while symmetry permits both linear and circular photon-drag responses at finite photon momentum. Nevertheless, despite these symmetry-allowed channels, the conventional photogalvanic and photon-drag effects vanish identically 
% we find that in both phases, the interband velocity matrix elements term (like $\mathbf{G}^{bc}_{{m,\textbf{k}_{+}};{n,\textbf{k}_{-}}} $ ) entering the conductivity tensors becomes identically zero.
% \textcolor{red}{Make this point more clear. Like how this quantity becomes zero in both phases}. 
Interestingly, the inclusion of Rashba spin--orbit coupling removes this trivial limit by inducing finite interband velocity matrix elements and nontrivial quantum-geometric quantities, thereby activating the symmetry-allowed photon-drag responses, whereas it still suppresses the conventional contribution. The resulting photon-drag conductivities for the  $p_x$- and $d_{x^2-y^2}$-wave magnets are discussed below.

%=================================================================
% \begin{figure}[b]
%     \includegraphics[width=\columnwidth]{p_wave_photondrag_im_shift_conductivity.pdf}
%     \caption{}
%     \label{fig:p_wave_im_sh}
% \end{figure}
%=================================================================

\subsection{$p_x$-wave unconventional magnet} \label{sec:$p_x$}
%~~~~~~~~~~~~~~~~~~~~~~~~~~~~~~~~~~~~~

%=================================================================
% \begin{figure}[t]
%     \includegraphics[width=\columnwidth]{d_wave_photondrag_RE_conductivity.pdf}
%     \caption{}
%     \label{fig:d_wave_real_inj}
% \end{figure}
%=================================================================

%=================================================================
% \begin{figure}[b]
%     \includegraphics[width=\columnwidth]{d_wave_photondrag_Im_conductivity.pdf}
%     \caption{}
%     \label{fig:d_wave_im_inj}
% \end{figure}
%=================================================================

The $p_x$-wave magnet preserves $\mathcal{T}$ while breaking $\mathcal{P}$, resulting in characteristic symmetry-governed components of the nonlinear optical response. In the presence of RSOC, the finite-photon-momentum photon-drag effect supports only linear injection and circular shift conductivities. Here, the linear response refers to the symmetric (real) part of the conductivity, whereas the circular response corresponds to the antisymmetric (imaginary) part. This behavior contrasts with the conventional photogalvanic effect ($q=0$), where circular injection and linear shift responses are only allowed \cite{Bristi_2026}. Although the general conductivity expressions in Eq.~(\ref{eq:gen_inj}) and Eq.~(~\ref{eq:gen_shift}) contain the total injection and shift contributions arising from both the conventional photogalvanic and photon-drag effects, the conventional photogalvanic responses in the $p_x$-wave magnet vanish due to the absence of Berry curvature and symplectic connection, rendering the second-order dc optical response purely photon-drag in nature. The finite photon momentum $q$ carried by the incident light effectively lowers the symmetry of the system, thereby enabling the emergence of the photon-drag effect. Accordingly, in this section, we focus on the linear injection and circular shift contributions to the photon-drag current.

\begin{figure*}[t]
    \includegraphics[width=0.8\textwidth]{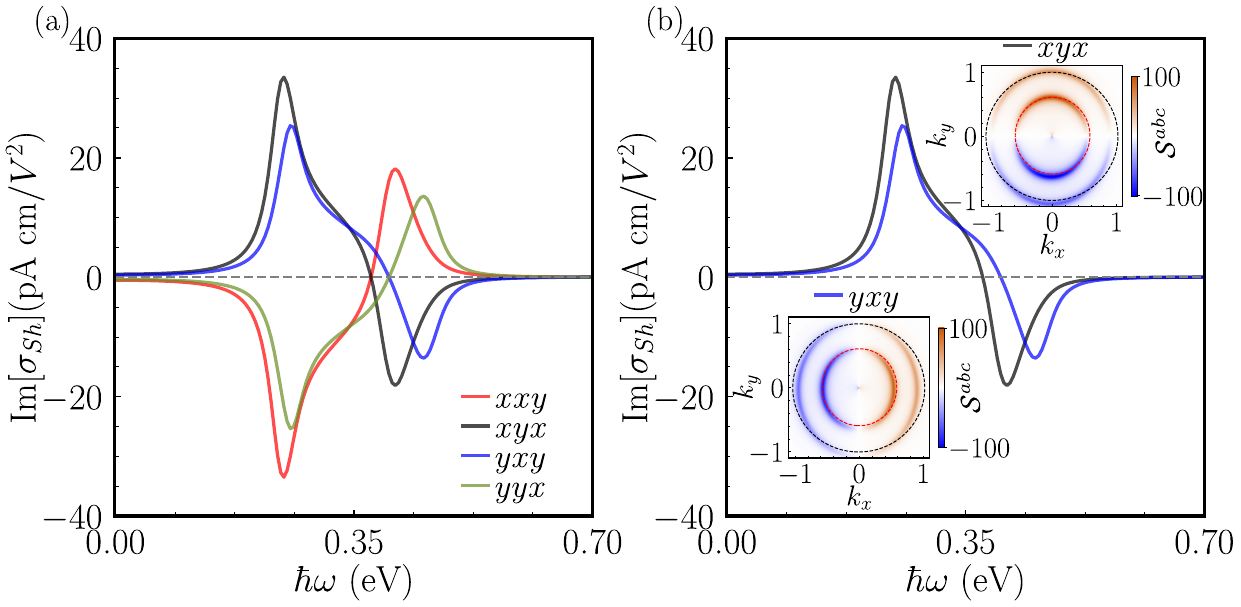}
    \caption{\textbf{Photon-drag shift response of the $p_x$-wave magnet with Rashba spin-orbit coupling:} (a) Frequency dependence of the imaginary part of the symmetry-allowed shift conductivity. The coexistence of time-reversal symmetry and broken inversion symmetry restricts the photon-drag shift response to its imaginary component, corresponding to the circular shift contribution. (b) Frequency dependence of $\mathrm{Im}[\sigma_{xyx}]$ and $\mathrm{Im}[\sigma_{yxy}]$, which display distinct spectral profiles and characteristic peak positions. The upper and lower insets depict the corresponding geometric transition weights $\mathcal{S}^{xyx}$ and $\mathcal{S}^{yxy}$ in the $k_x$--$k_y$ plane, respectively. The red and black dashed curves indicate the two Fermi surfaces. The contrasting distributions of $\mathcal{S}^{xyx}$ and $\mathcal{S}^{yxy}$ around the Fermi surfaces determine the relative spectral weights of the two tensor components and account for their different frequency-dependent behavior. Parameters are identical to those in Fig.~\ref{fig:p_wave_real_inj}.}
    \label{fig:p_wave_im_shift}
\end{figure*}

Since the $p_x$-wave magnet allows only a linear injection process in the presence of finite photon momentum  $\textbf{q}$, we first discuss the real part of the injection conductivity. Fig.~\ref{fig:p_wave_real_inj}(a) shows the frequency dependence of the various components of the real part of the injection current at a fixed chemical potential $\mu = 0.3$ eV. Owing to the intrinsic permutation symmetry of the injection conductivity with respect to the last two indices, the tensor components satisfy $\sigma_{yyx} = \sigma_{yxy}$ and $\sigma_{xxy} = \sigma_{xyx}$, since $\text{Re}[\mathbf{G}^{bc}_{{m,\textbf{k}_{+}};{n,\textbf{k}_{-}}}]=\text{Re}[\mathbf{G}^{cb}_{{m,\textbf{k}_{+}};{n,\textbf{k}_{-}}}]$. Consequently, only six independent tensor components remain, among which the $\sigma_{yyx}$ component is the most dominant. %Fig.~\ref{fig:p_wave_real_inj}(a) exhibits a divergence as $\omega \to 0$, originating from the $\omega^{-2}$ intraband contribution, followed by two prominent peaks associated with interband transitions. 
Fig.~\ref{fig:p_wave_real_inj}(a) exhibits two prominent peaks associated with interband transitions.
% These peaks arise from the presence of two concentric circular Fermi surfaces in the $p$-wave magnet: the momentum-space region between them remains Pauli unblocked, allowing optical transitions. Consequently, the two dominant interband transitions occur at the inner and outer edges of the Pauli-unblocked region. 
These two pronounced interband peaks originate from the two concentric circular Fermi surfaces of the $p_x$-wave magnet, as illustrated in the insets of Fig.~\ref{fig:p_wave_real_inj}(b).
% \textcolor{red}{The inner (red) and out/;'er (black) circles represent the valence- and conduction-band Fermi surfaces evaluated at the momentum-shifted states $\textbf{k}-\frac{\textbf{q}}{2}$ and $\textbf{k}+\frac{\textbf{q}}{2}$, respectively, which participate in the photon-drag transition.} 
The inner (red) and outer (black) circles correspond to the Fermi surfaces of the valence and conduction bands, respectively. Owing to the finite photon momentum $\mathbf{q}$, the optical transitions are nonvertical in momentum space, connecting the valence band state at $\mathbf{k}-\mathbf{q}/2$ to the conduction band state at $\mathbf{k}+\mathbf{q}/2$, and vice versa.
The momentum-space region enclosed between these two Fermi surfaces forms the Pauli-unblocked region, where interband optical transitions are allowed. Consequently, the two dominant interband transitions occur at the inner and outer edges of the Pauli-unblocked region. The color maps in the inset of Fig.~\ref{fig:p_wave_real_inj}(b) represent the momentum-resolved geometric transition weight entering the injection conductivity, which is defined by,
\begin{align*}
    &\mathcal{I}^{abc}=\\&-\!\! e\tau [v^a_{{m,\textbf{k}_{+}};{m,\textbf{k}_{+}}}\!\!-\!v^a_{{n,\textbf{k}_{-}};{n,\textbf{k}_{-}}}] \mathbf{G}^{bc}_{{m,\textbf{k}_{+}};{n,\textbf{k}_{-}}}\!\!\delta(\omega_{{m,\textbf{k}_{+}};{n,\textbf{k}_{-}}}-\omega).
\end{align*}

\noindent Although all tensor components exhibit contributions from both Fermi surfaces, the momentum-space distribution of the integrand differs significantly for different conductivity components because of the distinct angular dependence of the velocity-difference factor and the underlying geometric term. As a result, the lobes intersect the inner and outer Fermi surfaces at different regions of the $k_x$ and $k_y$ plane, leading to different relative weights of the two interband transitions. This difference in the momentum-space overlap gives rise to the distinct peak positions observed for the $\sigma_{xxx}$ and  $\sigma_{xyy}$ components. The same physical mechanism governs the remaining tensor components, although only these two representative cases are shown in the inset.
As the frequency increases, the $\sigma_{xxy}$ and $\sigma_{yxy}$ components exhibit a positive interband peak followed by a negative second peak, whereas the remaining components display the opposite behavior, namely an initial negative peak followed by a positive one. The emergence of this bipolar peak structure is fundamentally governed by the interplay between the altermagnetic band structure and the Fermi-Dirac distribution. In the absence of the Fermi function, the integrand of the injection conductivity remains positive throughout the Brillouin zone. However, once the Fermi function is incorporated, the integration becomes restricted to the Pauli-unblocked region lying between the two concentric Fermi surfaces. The Fermi-weighted contributions originating from the inner and outer boundaries of this region enter with opposite signs, leading to two resonant peaks of opposite polarity in the injection conductivity spectrum.\\

%====================================================================
We now examine the imaginary part of the shift conductivity, $\mathrm{Im}[\sigma_{\mathrm{Sh}}]$, in the $p_x$-wave altermagnet, as illustrated in Fig.~\ref{fig:p_wave_im_shift}(a). The figure presents the frequency dependence of the different tensor components of the imaginary shift response at a fixed chemical potential $\mu = 0.3$ eV. The shift conductivity is governed by the quantum-geometric connection-like quantity: $(\mathcal{C}^{abc}_{{m,\textbf{k}_{+}};{n,\textbf{k}_{-}}}-\mathcal{C}^{acb}_{{m,\textbf{k}_{+}};{n,\textbf{k}_{-}}}) $, whose real part contributes to $\mathrm{Im}[\sigma_{\mathrm{Sh}}]$. The real part of this quantity vanishes identically when the indices $b$ and $c$ are equal, leaving only four non-zero tensor components. Furthermore, the antisymmetric nature of this quantum geometric connection under index permutation implies that only two of these components are independent, satisfying the relations $\sigma_{yyx} = -\sigma_{yxy}$ and $\sigma_{xxy} = -\sigma_{xyx}$. As shown in Fig.~\ref{fig:p_wave_im_shift}(a), the component pairs $(\sigma_{xyx},\sigma_{xxy})$ and $(\sigma_{yyx},\sigma_{yxy})$ exhibit identical frequency dependence with opposite signs, although the latter pair possesses a comparatively smaller magnitude. 
% Similar to the injection response, all components exhibit a characteristic $\omega^{-2}$ divergence in the low-frequency regime due to dominant intraband processes, followed by pronounced interband resonances originating from transitions at the inner and outer boundaries of the Pauli-unblocked region between the two concentric Fermi surfaces.
Similar to the injection response, all components exhibit two pronounced interband resonances associated with transitions at the inner and outer boundaries of the Pauli-unblocked region between the two concentric Fermi surfaces, indicated by the black and red dashed lines in the insets of Fig.~\ref{fig:p_wave_im_shift}(b). The color maps show the momentum-resolved geometric transition weight $\mathcal{S}^{abc}$
 entering the shift conductivity, which is defined by
\begin{equation*}
    \mathcal{S}^{abc}=  ie
\left[\mathcal{C}^{abc}_{{m,\textbf{k}_{+}};{n,\textbf{k}_{-}}}-\mathcal{C}^{acb}_{{m,\textbf{k}_{+}};{n,\textbf{k}_{-}}}\right]\delta(\omega_{{m,\textbf{k}_{+}};{n,\textbf{k}_{-}}}-\omega).
\end{equation*}
 
\noindent Similar to the injection transition weight $\mathcal{I}^{abc}$, $\mathcal{S}^{abc}$ exhibits distinct lobe structures for different tensor components. These lobes overlap with the inner and outer Fermi-surface boundaries differently along the $k_x$ and $k_y$ directions, leading to component-dependent contributions from the two boundaries and, consequently, distinct resonance positions and amplitudes in the shift conductivity. A defining feature of the interband response is the bipolar peak structure, where each component undergoes a sharp sign reversal. Specifically, $\sigma_{xyx}$ and $\sigma_{xxy}$ exhibit an initial negative peak at the inner boundary of the unblocked region, followed by a positive resonance at the outer boundary. Conversely, the $\sigma_{yyx}$ and $\sigma_{yxy}$ components display an inverted polarity. This behavior highlights the role of the Fermi-Dirac distribution as a spectral filter, as the integrand with the Fermi function factor restricts transitions to a momentum-space window where the contributions from the inner and outer Fermi-surface edges enter with opposite signs, like the injection current.

\begin{figure}[t]
    \includegraphics[width=\columnwidth]{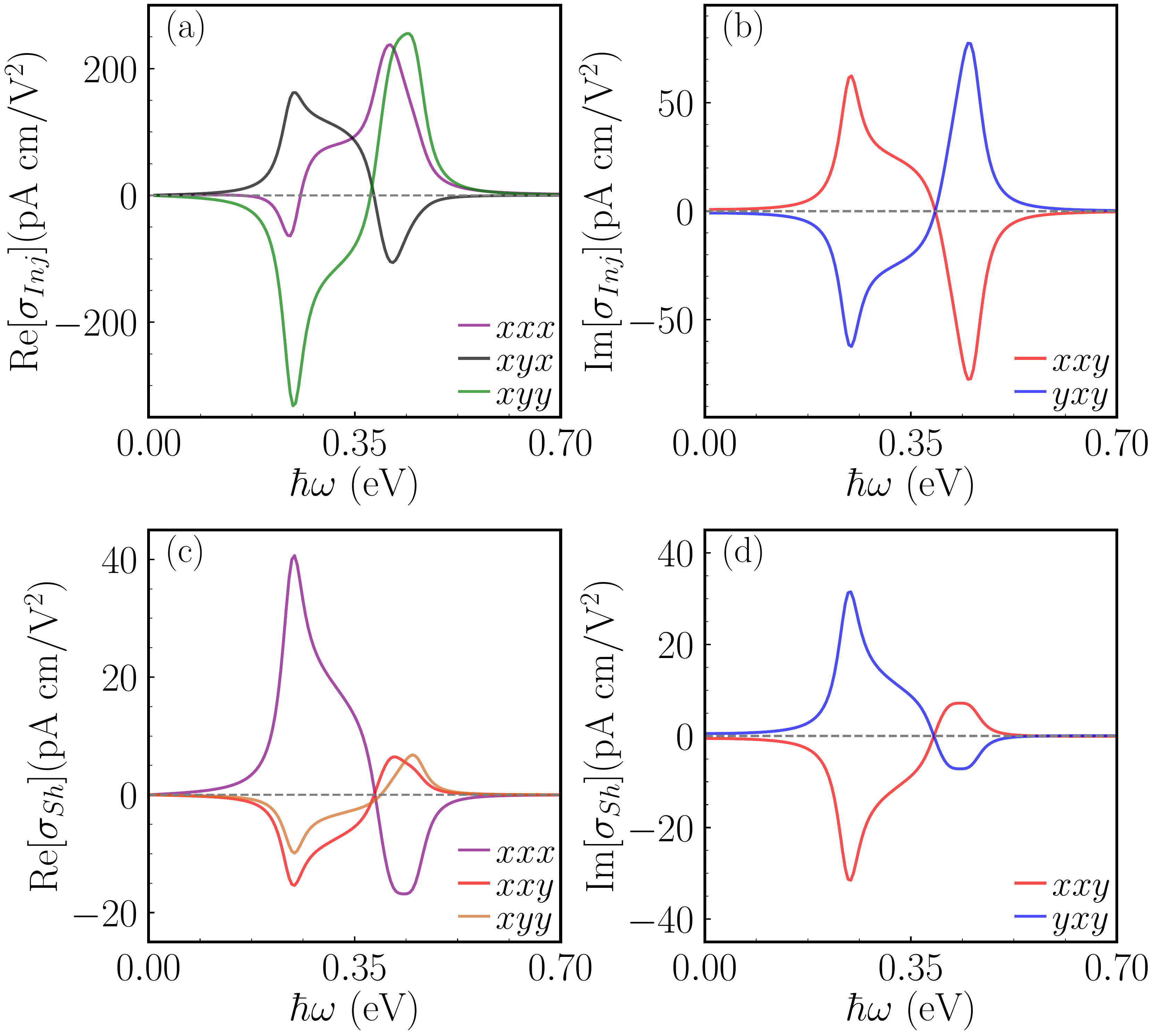}
    \caption{\textbf{Photon-drag response of the $d$-wave magnet with Rashba spin-orbit coupling:} Frequency dependence of the symmetry-allowed independent components of the (a) real part of the injection conductivity, (b) imaginary part of the injection conductivity, (c) real part of the shift conductivity, and (d) imaginary part of the shift conductivity. The simultaneous breaking of inversion and time-reversal symmetries in the $d$-wave magnet permits both the real (linear) and imaginary (circular) components of the injection and shift conductivities to contribute to the photon-drag response. The model parameters used are ($t$, $J$, $\lambda$, $\tau$, $T$, $q_x$, $q_y$) = ($0.5~\mathrm{eV.nm}^2$, $0.1~\mathrm{eV.nm}$, $0.2~\mathrm{eV.nm}$, $33~\mathrm{fs}$, $10~\mathrm{K}$, $0.01~\mathrm{nm}^{-1}$, $0.01~\mathrm{nm}^{-1}$).}
    \label{fig:d_wave_photondrag}
\end{figure}
%~~~~~~~~~~~~~~~~~~~~~~~~~~~~~~~~~~~~~
\subsection{$d_{x^2-y^2}$-wave magnet} \label{sec:$d_wave$}
%~~~~~~~~~~~~~~~~~~~~~~~~~~~~~~~~~~~~~

We now extend our investigation to the photon-drag effect in the $d$-wave altermagnet. Unlike the $p_x$-wave case, the $d_{x^2-y^2}$-wave magnet, in the presence of RSOC, simultaneously breaks both $\mathcal{T}$ and $\mathcal{P}$. Consequently, both linear and circular components of the injection and shift conductivities become finite at finite photon momentum. Similar to the $p$-wave magnet, the symmetry of the $d$-wave state constrains the total conductivity to vanish in the $q=0$ limit upon integration over the Brillouin zone, rendering the nonlinear dc optical response purely photon-drag in nature. In the following analysis, we provide a comprehensive study of both the linear and circular photon-drag contributions, highlighting how the intrinsic $d$-wave anisotropy shapes the spectral characteristics of these nonlinear currents.\\

The frequency dependence of the various components of the real part of the injection conductivity is shown in Fig.~\ref{fig:d_wave_photondrag}(a), which exhibits behavior similar to the $p_x$-wave case. As the frequency increases, the $\sigma_{xyx}$ component displays a positive interband peak followed by a negative second peak, whereas the remaining components exhibit the opposite trend, with an initial negative peak followed by a positive one. Due to the combined symmetries $\mathcal{C}_{4z}\mathcal{T}$ and $\mathcal{M}_x\mathcal{T}$, the tensor elements obey $\sigma_{xxx}=\sigma_{yyy}$, $\sigma_{xyy}=\sigma_{yxx}$, and $\sigma_{xyx}=\sigma_{xxy}=\sigma_{yyx}=\sigma_{yxy}$, reducing the number of independent components to three, with $\sigma_{xyy}$ being the most prominent. The imaginary part of the injection conductivity in the $d_{x^2-y^2}$-wave magnet arises from circular injection processes enabled by broken time-reversal symmetry. As the underlying quantum geometry factor governing this response is $\text{Im}[\mathbf{G}^{bc}_{{m,\textbf{k}_{+}};{n,\textbf{k}_{-}}} ]$, which is antisymmetric under the exchange of the last two optical indices, only the $\sigma_{xxy}$ and $\sigma_{yxy}$ components are nonzero. As shown in Fig.~\ref{fig:d_wave_photondrag}(b), their frequency dependence follows the same trend as observed for the real part, with the two components being equal in magnitude but opposite in sign. \\

%=================================================================
% \begin{figure}[t]
%     \includegraphics[width=\columnwidth]{d_wave_photondrag_im_shift_conductivity.pdf}
%     \caption{}
%     \label{fig:d_wave_im_shift}
% % \end{figure}
%=================================================================

Likewise, the photon-drag shift response of the $d_{x^2-y^2}$-wave magnet allows both linear and circular shift conductivities. The frequency dependence of the independent components of the real part of the shift conductivity at a fixed chemical potential $\mu = 0.3$ eV is presented in Fig.~\ref{fig:d_wave_photondrag}(c). Owing to the combined symmetries $\mathcal{C}_{4z}\mathcal{T}$ and $\mathcal{M}_x\mathcal{T}$, the tensor components satisfy $\sigma_{xxx}=\sigma_{yyy}$, $\sigma_{xyy}=\sigma_{yxx}$, and $\sigma_{xyx}=\sigma_{xxy}=\sigma_{yyx}=\sigma_{yxy}$, leaving only three independent components, among which $\sigma_{xxx}$ exhibits the dominant contribution. Similar to the behavior discussed for the $p_x$-wave magnet, the shift response displays two pronounced interband bipolar peaks arising from optical transitions at the boundaries of the Pauli-unblocked region. The imaginary part of the shift conductivity in the $d_{x^2-y^2}$-wave magnet originates from the circular photon-drag process, which is governed by the antisymmetric quantum-geometric structure. Similar to the $p_x$-wave case, the underlying antisymmetric quantum-geometric quantity $(\mathcal{C}^{abc}_{{m,\textbf{k}_{+}};{n,\textbf{k}_{-}}}-\mathcal{C}^{acb}_{{m,\textbf{k}_{+}};{n,\textbf{k}_{-}}})$, allows only four finite tensor components, of which only two are independent, satisfying the relations $\sigma_{yyx} = -\sigma_{yxy}$ and $\sigma_{xxy} = -\sigma_{xyx}$. As shown in Fig.~\ref{fig:d_wave_photondrag}(d), the frequency dependence of these components follows the same behavior as the real part of the shift response, with the two independent components remaining equal in magnitude but opposite in sign.

To estimate the experimentally relevant magnitude of the injection and shift currents, we consider RuO$_2$ as a representative $d_{x^2-y^2}$-wave altermagnetic candidate~\cite{Andriy_2024,Liu_2024,plouff_2025}. Using the parameters $t=1\,\mathrm{eV}$, $J=0.5\,\mathrm{eV}$, $\lambda=0.4\,\mathrm{eV}$, and $\mu=0.3\,\mathrm{eV}$~\cite{Libor_2020}, together with a relaxation time of $\tau=0.016\,\mathrm{ps}$ at $T=10\,\mathrm{K}$, we estimate the photon-drag injection and shift conductivities under a THz-frequency optical pump. The resulting conductivities are of order $20$--$100\,\mathrm{pA\,cm/V^2}$, indicating that the predicted photon-drag response should be experimentally accessible.

%\textcolor{red}{We can think of doing a tight-binding calculation and calculate the extrinsic quadrupolar contribution.}

%
\section{Summary and Conclusion}
\label{sec: conclusion}

In this work, we systematically investigated the geometric quantity-driven photon-drag effect in $p_x$-wave and $d_{x^2-y^2}$-wave magnetic systems in the presence of RSOC. We demonstrated that the finite-photon-momentum-mediated photon-drag effect generates a finite dc optical response even when the conventional photogalvanic response vanishes identically in these systems. In this context, the finite-momentum photon-drag effect provides a particularly important mechanism for realizing measurable nonlinear optical responses without the need for external magnetic fields or additional symmetry-breaking perturbations. We also find that these two unconventional magnetic systems exhibit qualitatively distinct nonlinear responses dictated by their symmetry properties. The $p_x$-wave magnet, which preserves $\mathcal{T}$ while breaking $\mathcal{P}$, supports only linear injection and circular shift conductivities, whereas the $d_{x^2-y^2}$-wave altermagnet, which simultaneously breaks both symmetries, admits all linear and circular components of the injection and shift responses. The bipolar peak structures and sign reversals in the conductivity spectra emerge from the selective momentum-space filtering imposed by the band structure and the Fermi-Dirac distribution. Furthermore, the nonlinear conductivity tensors exhibit strong anisotropy due to symmetry-imposed constraints on the underlying quantum-geometric matrix elements. In particular, the combined symmetries $C_{4z}\mathcal{T}$ and $M_x\mathcal{T}$ in the $d$-wave altermagnet substantially reduce the number of independent tensor components, providing clear optical fingerprints of the underlying altermagnetic order.

%The calculated conductivity spectra display characteristic low-frequency divergences originating from intraband processes, followed by pronounced interband resonances near the boundaries of the Pauli-unblocked region. 

Experimentally, the photon-drag response can be identified through its dependence on photon momentum and light polarization. Reversing the incidence angle, $q\rightarrow -q$, isolates the component of the photocurrent that is odd in $q$~\cite{xie_pnas2025, Strait_prb2019, Shalygin_jpcs2017}, while polarization-resolved measurements can be used to distinguish the symmetry-allowed injection and shift contributions~\cite{Shi_prl2021, xie_pnas2025}. We note that a linear-in-q photon-drag contribution arising from the electric-quadrupole correction has recently been proposed, with this contribution being connected to the multiband quantum metric tensor~\cite{gassner_arxiv2026}. Such a quadrupolar contribution, however, is relevant only for systems with three or more bands. Since the minimal model considered in this work is restricted to two bands, we neglect this contribution here. Extending the present analysis to a multiband tight-binding model would therefore be an interesting direction for future work, allowing the role of the electric-quadrupole contribution to be investigated explicitly. In addition, disorder-mediated mechanisms, such as side-jump and skew-scattering contributions, may also contribute to the photon-drag response. A systematic investigation of these effects is beyond the scope of the present work and is left for future studies. 

Overall, our results establish unconventional magnets as an ideal platform for realizing and probing pure geometric photon-drag transport phenomena. Furthermore, our predictions can be directly tested experimentally in $d$-wave altermagnetic materials such as RuO$_2$~\cite{libor_prx2022}, MnF$_2$~\cite{Bhowal_2024}, and MnTe~\cite{Mazin_2023}, as well as in candidate $p$-wave magnetic systems including CeNiAsO~\cite{hellenes2023} and NiI$_2$~\cite{cuono2026}. The strong sensitivity of the nonlinear response to altermagnetic symmetry, finite photon momentum, and quantum geometry demonstrates that photon-drag spectroscopy can serve as a powerful probe of momentum-dependent spin textures and symmetry-breaking mechanisms in unconventional magnetic systems, opening new possibilities for symmetry-controlled nonlinear optoelectronic applications in quantum magnetic materials.

\section*{Acknowledgment}

S.~N. acknowledges financial support from the Anusandhan National Research Foundation (ANRF), Government of India via the Prime Minister's Early Career Research Grant: ANRF/ECRG/2024/005947/PMS. P. B. acknowledges the financial support by the Anusandhan National Research Foundation SERB SURE under project number SUR/2022/000289.  B.~G. is supported by INSPIRE, DST, Government of India.
%

% \onecolumngrid
% \appendix
% \section{Derivation of Equations} \label{Appendix:A}

% %
% %
% \twocolumngrid

\bibliography{ref}

\end{document}